\documentclass[10pt,letterpaper]{article}
\usepackage[top=0.85in,left=2.75in,footskip=0.75in]{geometry}

\usepackage{amsmath,amssymb,mathtools}

\usepackage{changepage}

\usepackage[utf8x]{inputenc}

\usepackage{textcomp,marvosym}

\usepackage{cite}

\usepackage{nameref,hyperref}

\usepackage[right]{lineno}

\usepackage{microtype}
\DisableLigatures[f]{encoding = *, family = * }

\usepackage[table]{xcolor}

\usepackage{array}

\newcolumntype{+}{!{\vrule width 2pt}}

\newlength\savedwidth

\usepackage{setspace} 
\raggedright
\usepackage[aboveskip=1pt,labelfont=bf,labelsep=period,justification=raggedright,singlelinecheck=off]{caption}

\makeatletter
\renewcommand{\@biblabel}[1]{\quad#1.}
\makeatother

\usepackage{lastpage,fancyhdr,graphicx}
\usepackage{epstopdf}
\fancyheadoffset[L]{2.25in}
\fancyfootoffset[L]{2.25in}
\begin{document}
\vspace*{0.2in}

\begin{flushleft}
{\Large
\textbf\newline{Thermodynamic validity criterion for the irreversible Michaelis-Menten equation} 
}
\newline
\\
Valérie Voorsluijs\textsuperscript{1,2*},
Francesco Avanzini\textsuperscript{1**},
Massimiliano Esposito\textsuperscript{1***}
\\
\bigskip
\textbf{1} Complex Systems and Statistical Mechanics, Department of Physics and Materials Science, University of Luxembourg, L-1511 Luxembourg, Luxembourg.
\\
\textbf{2} Luxembourg Centre for Systems Biomedicine, University of Luxembourg, L-4365 Esch-sur-Alzette, Luxembourg.
\\
\bigskip

* valerie.voorsluijs@uni.lu\\
** francesco.avanzini@uni.lu\\
*** massimiliano.esposito@uni.lu

\end{flushleft}
\section*{Abstract}
Enzyme kinetics is very often characterised by the irreversible Michaelis-Menten (MM) equation. However, in open chemical reaction networks such as metabolic pathways, this approach can lead to significant kinetic and thermodynamic inconsistencies. Based on recent developments in nonequilibrium chemical thermodynamics, we present a validity criterion solely expressed in terms of the equilibrium constant of the enzyme-catalysed reaction. When satisfied, it guarantees the ability of the irreversible MM equation to generate kinetic and thermodynamic data that are quantitatively reliable for reasonable ranges of concentrations. Our validity criterion is thus a precious tool to ensure reliable kinetic and thermodynamic modelling of pathways. We also show that it correctly identifies the so-called irreversible enzymatic reactions in glycolysis and Krebs cycle. 

\section*{Author summary}
Living systems need to continuously feed on nutrients present in their surrounding to gather the energy necessary to their survival. That energy is extracted by a complex network of chemical reactions which constitutes metabolism. Enzymes are proteins that speed up and regulate the reactions without being consumed or produced by them. Most enzymes are characterised by only two kinetic parameters which are measured experimentally and reported in tables. By doing so, one implicitly assumes enzymes follow a kinetic mechanism called the irreversible Michaelis-Menten (MM) scheme. While this assumption usually holds \textit{in vitro}, it is by no means ensured to apply to living systems. Furthermore, the irreversible MM scheme is \textit{a priori} inconsistent with thermodynamics, the theory describing energy conversion in systems ranging from car engines to molecular motors. In this paper, we propose a way to easily test if the MM equation holds and explain how to make the irreversible MM scheme consistent with thermodynamics. Beside providing precious information about which enzymes need to be characterised beyond the irreversible MM scheme, our work paves the way for more realistic thermodynamic considerations in biochemistry to understand the efficiency with which living systems process energy. 

\section*{Introduction}
Enzymes are ubiquitous proteins catalysing biochemical reactions with very high efficiency. They are involved in a broad array of physiological processes, ranging from metabolism to cell signalling, and usually display a high selectivity with respect to their substrate. Enzyme kinetics can be subject to regulation by activators and/or inhibitors, and the fluxes along the biochemical pathways can hence be modulated in the course of time and, for example, adapt to the metabolic demand. Variations in kinetic rates can also indicate the presence of a disease or be induced by a pharmacological treatment. Appropriate mathematical descriptions of enzyme kinetics are thus crucial to improve the current understanding of physiology in health and disease.

The analysis of complex reactions networks such as biochemical pathways usually relies on reduced mathematical models incorporating a smaller number of variables and/or parameters~\cite{murray_mathematical_2002, keener_mathematical_2009}. Simplifications can be achieved by nondimensionalisation and scaling of the governing equations~\cite{lin_mathematics_1988}, and/or by making assumptions regarding the kinetics. More precisely, when the variables of the system evolve on very different timescales, the dynamics is dominated by the behaviour of the slow variables and the fast variables, which are driven towards a pre-equilibrium or a steady state, can be adiabatically eliminated~\cite{van_kampen_elimination_1985}. This idea is at the core of the equilibrium and quasi-steady-state approximations~\cite{murray_mathematical_2002, keener_mathematical_2009}. However, this elimination can lead to thermodynamic inconsistencies in the model~\cite{shiner_fluxes_1981} and these aspects should be taken into account~\cite{wachtel_thermodynamically_2018}.

In the context of enzyme kinetics, the Michaelis-Menten (MM) equation is extensively used to characterise the rate of enzyme-catalysed reactions~\cite{beard_chemical_2008,cornish-bowden_fundamentals_2013}. This equation not only relies on the steady-state approximation, but additionally assumes an \textit{irreversible} mechanism. While some enzymes such as invertases, phosphatases and peptidases can indeed be considered as ``one-way" catalysts, where the backward transformation is negligible, most enzymatic reactions are reversible~\cite{cornish-bowden_fundamentals_2013,bisswanger_enzyme_2017}. If the reaction product has not accumulated in sufficiently large quantities to make the reverse reaction significant, as is the case at the beginning of the reaction, the irreversible MM equation can be applied to reversible processes without jeopardising the relevance of kinetic results~\cite{bisswanger_enzyme_2017}. For example, enzyme characterisation typically takes place in this regime in order to avoid the effects of the reverse reaction~\cite{cornish-bowden_fundamentals_2013}. However, how can we ascertain that the MM equation provides a good description of the product formation rate in reaction pathways~\cite{curien_understanding_2009}? In this case, each reaction can be assimilated to an open system where the substrate and product are continuously injected and removed, which contrasts with the conditions of enzyme characterisation, usually proceeding in closed environments (\textit{i.e}.\ without exchange of matter).

In order to provide a more robust framework for enzyme modelling, it thus appears crucial to derive explicit conditions of validity for the irreversible MM equation. Different theoretical approaches have been adopted to this end~\cite{min_when_2006, kolomeisky_michaelismenten_2011, cao_michaelismenten_2011}, but mainly focussed on kinetic and conformational aspects of catalysis. So far, the thermodynamic validity of the MM equation has been side\-stepped, probably because of the theoretical issues raising from the irreversible character of the reaction~\cite{gorban_thermodynamics_2013}. In this paper, we address that gap by deriving a \textit{nonequilibrium thermodynamic criterion} indicating whether the irreversible MM equation is a valid model for a given single-substrate enzymatic reaction. The novelty of our approach is twofold. We not only provide modellers with an explicit criterion indicating when the irreversible MM rate is thermodynamically consistent. We also show that nonequilibrium thermodynamics can be applied to irreversible reactions.

This paper is organised as follows. In ~\nameref{section:methods}, we describe and compare the irreversible and reversible MM equations before introducing the entropy production rate (EPR) of reactive systems. This quantity is used to derive our validity criterion, which is one of the key results of this paper. The validity of our approach is examined in two ways. On the one hand, we test our criterion by applying it to well-characterised biochemical reactions. On the other hand, we quantify the error made on the EPR when it is estimated \textit{via} the irreversible flux. We finally summarise our results and discuss the relevance of our approach to less well characterised pathways.

\section*{Methods}
\label{section:methods}
\subsection*{The Michaelis-Menten equations}
\label{section_MM}

\subsubsection*{Irreversible case}
The most common mechanism representing an enzyme-catalysed reaction of the type $\mathrm{S} \rightarrow \mathrm{P}$ follows the kinetic scheme:
\begin{equation}
\mathrm{E} + \mathrm{S} \xrightleftharpoons[k_{-1}]{k_1} \mathrm{ES} \xrightarrow{k_2} \mathrm{E} + \mathrm{P},
\label{eq_scheme_irr_MM}
\end{equation}
where substrate S binds to enzyme E to form a complex ES that further releases the intact enzyme and the reaction product P. This mechanism, which neglects the reverse reaction and regulation processes such as product inhibition, was used to derive the irreversible MM equation:
\begin{equation}
J_\mathrm{irr}=\frac{k_{cat} \left[\mathrm{E}\right]_0 \left[\mathrm{S}\right]}{\left[\mathrm{S}\right]+K_M},
\label{eq_general_MM}
\end{equation}
where $J_\mathrm{irr}$ is the reaction rate, $\left[\mathrm{S}\right]$ is the concentration of the substrate, $\left[\mathrm{E}\right]_0$ is the total concentration of the enzyme ($\left[\mathrm{E}\right]_0=\left[\mathrm{E}\right]+\left[\mathrm{ES}\right]$), $k_{cat}$ is the catalytic constant or turnover number, and $K_M$ is the Michaelis constant \cite{cornish-bowden_fundamentals_2013}. As ES is initially absent, $\left[\mathrm{E}\right]_0$ also corresponds to the initial concentration of E. For the two-step mechanism shown in \eqref{eq_scheme_irr_MM}, $k_{cat}=k_2$ and $K_M=\frac{k_{-1}+k_2}{k_1}$, but Eq~\eqref{eq_general_MM} also applies to more complex kinetic schemes involving multiple steps.
The catalytic efficiency of an enzyme, defined as $k_{cat}/K_M$, and the limiting rate, $V=k_{cat} \left[\mathrm{E}\right]_0$, are also widely used in the literature to characterise enzymes.

The derivation of Eq~\eqref{eq_general_MM} relies on the assumption that the formation of the complex is fast and the enzyme rapidly saturated, leading to a quasi-steady state for the complex (\textit{i.e.} $d_t \left[\mathrm{ES}\right]=0$) \cite{segel_quasi-steady-state_1989, murray_mathematical_2002, keener_mathematical_2009}. This condition is fulfilled if the substrate is in large excess with respect to the enzyme or if $K_M$ is relatively low compared to the substrate concentration. A deeper timescale analysis of the system provides a refined condition guaranteeing the validity of the quasi-steady-state approximation \cite{murray_mathematical_2002}:
\begin{equation}
\frac{\left[\mathrm{E}\right]_0}{\left[\mathrm{S}\right]_0+K_M}\frac{1}{1+\left(k_{-1}/k_2\right)+\left(k_1 \left[\mathrm{S}\right]_0 / k_2\right)} \ll 1,
\end{equation}
where $\left[\mathrm{S}\right]_0$ is the initial substrate concentration.

\subsubsection*{Reversible case}
Using the quasi-steady-state approximation for ES, a rate expression can be derived similarly for the reversible scheme
\begin{equation}
\mathrm{E} + \mathrm{S} \xrightleftharpoons[k_{-1}]{k_1} \mathrm{ES} \xrightleftharpoons[k_{-2}]{k_2} \mathrm{E} + \mathrm{P},
\label{eq_scheme_rev_MM}
\end{equation}
and reads
\begin{equation}
J_\mathrm{rev}=\frac{\left[\mathrm{E}\right]_0 \left(k_s \left[\mathrm{S}\right] - k_p \left[\mathrm{P}\right]\right)}{1 + \frac{\left[\mathrm{S}\right]}{K_{Ms}} + \frac{\left[\mathrm{P}\right]}{K_{Mp}}},
\label{eq_rev_MM}
\end{equation}
where $\left[\mathrm{P}\right]$ is the product concentration, $k_s=\frac{k_1 k_2}{k_{-1}+k_2}$, $k_p=\frac{k_{-1} k_{-2}}{k_{-1}+k_2}$, $K_{Ms}=\frac{k_{-1}+k_2}{k_1}$, $K_{Mp}=\frac{k_{-1}+k_2}{k_{-2}}$ and the other notations are the same as in Eq~\eqref{eq_general_MM}. Although feasible and based on experimental procedures similar to the handling of irreversible enzymes, the characterisation of reversible enzymatic processes is less straightforward, since it requires measurements for backward and forward reactions \cite{lee_kinetics_2000,bisswanger_enzyme_2017}, and tends to be avoided. Using the irreversible MM equation is thus usually more convenient and we now examine the thermodynamic conditions under which this approximation can be made.

\subsection*{Nonequilibrium thermodynamics of reversible processes}
\label{section_NET}
To investigate the validity of the irreversible MM equation, we base our analysis on the EPR, $\dot{\Sigma}$.
As a consequence of the second law of thermodynamics, the EPR of a system is always non-negative and is equal to zero at equilibrium. In this Section, we introduce key thermodynamic quantitites involved in the computation of EPR for elementary processes and the reversible MM scheme, before extending this formalism to the irreversible MM scheme, as further detailed in \nameref{section:results}.
A more systematic description of the nonequilibrium thermodynamics of chemical reaction networks can be found in refs \cite{rao_nonequilibrium_2016} and \cite{wachtel_thermodynamically_2018}.

\subsubsection*{Elementary processes}
In the absence of mass and heat transport, the EPR associated with a set of $N$ chemical reactions is given by
\begin{equation}
\dot{\Sigma}=\sum_{\rho=1}^N  J_{\rho}\, \frac{\mathcal{A}_{\rho}}{T} \geq 0, 
\label{eq_EPR}
\end{equation}
where $T$ is the absolute temperature while $J_\rho$ and $\mathcal{A}_{\rho}$ are the net rate and affinity of reaction $\rho$, respectively. In nonequilibrium thermodynamics, the latter factor is often referred to as the \textit{force} acting on the system while the first factor is interpreted as the response or the \textit{flux} of the system. The equilibrium state is characterised by zero forces and hence zero fluxes.

For elementary processes, the net rate is $J_\rho=J_{+\rho}-J_{-\rho}$ and, according to the \textit{law of mass action}, the reaction rates associated with the forward ($+\rho$) and backward ($-\rho$) reactions are given by
\begin{equation}
J_{\pm\rho}=k_{\pm\rho }\prod_{\sigma} Z_{\sigma}^{\nu_{\sigma}^{\pm\rho}},
\label{eq_flux}
\end{equation}
where $Z_{\sigma}$ is the concentration of species $\sigma$ and $\nu_{\sigma}^{\pm\rho}$ is the stoichiometric coefficient of species $\sigma$ in reaction $\pm\rho$. The forward and backward reaction currents become equal at equilibrium (this is the principle of \textit{detailed balance}), so the net reaction rate becomes zero. It follows that
\begin{equation}
\frac{k_{+\rho}}{k_{-\rho}}= \prod_{\sigma} Z^{\mathrm{eq}\, \mathbb{S}_\sigma^\rho}_{\sigma},
\label{eq_detailed_balance}
\end{equation}
where $\mathbb{S}_\sigma^\rho=\nu_{\sigma}^{-\rho}-\nu_{\sigma}^{+\rho}$ is the net stoichiometric coefficient of species $\sigma$ in reaction $\rho$ and superscript ``eq'' denotes equilibrium conditions. The right-hand side of Eq~\eqref{eq_detailed_balance} is the equilibrium constant, $K_{\rho}$, related to the standard Gibbs free energy of reaction by
\begin{equation}
K_{\rho}=\exp\left({-\frac{\Delta_\rho G^\circ}{R T}}\right)=\frac{k_{+\rho}}{k_{-\rho}},
\label{eq_equilibrium_constant}
\end{equation}
where $R$ is the gas constant.

On the other hand, the affinity of reaction $\rho$ is defined by \cite{de_groot_non-equilibrium_1984}
\begin{equation}
\mathcal{A}_{\rho}=- \sum\limits_{\sigma} \mathbb{S}_\sigma^{\rho} \mu_{\sigma},
\label{eq_affinity}
\end{equation}
where $\mu_{\sigma}$ is the chemical potential of species $\sigma$. Under the hypothesis of \textit{local equilibrium}, \textit{i.e}.\ state variables such as temperature and pressure relax to equilibrium on a much faster timescale than reaction rates so the expressions for the thermodynamic potentials derived at equilibrium still hold locally out-of-equilibrium \cite{de_groot_non-equilibrium_1984}, the chemical potential $\mu_{\sigma}$ is given by
\begin{equation}
\mu_{\sigma}=\mu_{\sigma}^\circ + R T \ln Z_{\sigma},
\label{eq_chemical_potential}
\end{equation}
where $\mu_{\sigma}^\circ$ denotes the standard chemical potential of species $\sigma$. Standard conditions correspond to atmospheric pressure $p^\circ=1$ bar and molar concentrations $Z_{\sigma}^\circ=1\,\mathrm{mol\,L}^{-1}$. Finally, standard chemical potentials are directly related to the standard Gibbs free energy of reaction
\begin{equation}
\Delta_\rho G^\circ=\sum\limits_{\sigma} \mathbb{S}_{\sigma}^{\rho} \mu_{\sigma}^\circ.
\label{eq_deltaG0}
\end{equation}

Combining equations \eqref{eq_flux}, \eqref{eq_equilibrium_constant}, \eqref{eq_affinity}, \eqref{eq_chemical_potential} and \eqref{eq_deltaG0} yields
\begin{equation}
\mathcal{A}_{\rho}= R T \ln \frac{k_{+\rho} \prod\limits_{\sigma} Z_{\sigma}^{\nu^{+\rho}_{\sigma}}}{k_{-\rho} \prod\limits_{\sigma} Z_{\sigma}^{\nu^{-\rho}_{\sigma}}}=RT \ln \frac{J_{+\rho}}{J_{-\rho}}
\label{eq_affinity_elem}
\end{equation}
and the EPR, which can then be rewritten as
\begin{equation}
\dot{\Sigma}=\sum\limits_{\rho}\left(J_{+\rho}-J_{-\rho}\right) R \ln \frac{J_{+\rho}}{J_{-\rho}}
\label{eq_EPR_elem}
\end{equation}
is thus necessarily non-negative. Also, it clearly appears in Eq~\eqref{eq_EPR_elem} that the affinity and hence the EPR diverge if process $\rho$ is irreversible, \textit{i.e.} $J_{-\rho} \rightarrow 0$.

\subsubsection*{Reversible MM scheme}
In living systems, biochemical reactions are maintained out of equilibrium due to exchanges with the environment (influx of nutrients, excretion, \textit{etc}), which influences the steady-state of the system or even leads to more complex dynamics such as oscillations \cite{murray_mathematical_2002,keener_mathematical_2009}. To mimic this feature at the scale of a given pathway or catalytic step, it is usual to assume that the input and output chemicals are \textit{chemostatted species}, \textit{i.e.} their concentration is maintained constant in the course of time due to continuous exchanges with the environment and homeostasis. In the MM scheme, we thus consider that S and P are chemostatted, with concentrations $\left[\mathrm{S}\right]$ and $\left[\mathrm{P}\right]$, respectively. By doing so, we assume that the dynamics of the global pathway has relaxed to a steady state or oscillates at a frequency much slower than the timescale of the enzymatic reaction of interest.

At steady state (denoted ``ss"), Eq~\eqref{eq_EPR_elem} is equivalent to
\begin{equation}
\dot{\Sigma}^\mathrm{ss}_\mathrm{rev}=J_\mathrm{rev}\frac{\mathcal{A}_\mathrm{rev}}{T},
\label{eq_EPR_rev_gen}
\end{equation}
where $J_\mathrm{rev}=J_1^\mathrm{ss}=J_2^\mathrm{ss}$ is given by Eq~\eqref{eq_rev_MM} and $\mathcal{A}_\mathrm{rev}=\mathcal{A}_1 + \mathcal{A}_2=RT\ln\frac{k_1 k_2 \left[\mathrm{S}\right]}{k_{-1} k_{-2} \left[\mathrm{P}\right]}$, with indexes 1 and 2 referring to the first and second elementary steps constituting the reversible MM scheme \eqref{eq_scheme_rev_MM}.

Put in this form, the affinity and the EPR diverge in the irreversible limit, which corresponds to $k_{-2} \left[\mathrm{P}\right] \rightarrow 0$. However, the affinity can be rewritten using the definition of chemical potential (Eq~\eqref{eq_chemical_potential}). We then have
\begin{equation}
\mathcal{A}_\mathrm{rev}=\mu_{\mathrm{S}}^\circ-\mu_{\mathrm{P}}^\circ + R T \ln \frac{\left[\mathrm{S}\right]}{\left[\mathrm{P}\right]},
\label{eq_affinity_chemical_potentials}
\end{equation}
where $\mu_{\mathrm{P}}^\circ-\mu_{\mathrm{S}}^\circ=\Delta_{\mathrm{rev}} G^\circ$ is the standard Gibbs free energy of the reaction $\mathrm{S} \rightarrow \mathrm{P}$. It can be expressed as $\Delta_{\mathrm{rev}} G^\circ=\Delta_f G^\circ_\mathrm{P}-\Delta_f G^\circ_\mathrm{S}$, where $\Delta_f G^\circ_\mathrm{P}$ and $\Delta_f G^\circ_\mathrm{S}$ are the standard Gibbs free energies of formation of P and S, respectively. These quantities are thermodynamic data usually available in tables and $\mathcal{A}_\mathrm{rev}$ can thus be calculated for positive and non-zero values of $\left[\mathrm{S}\right]$ and $\left[\mathrm{P}\right]$, as is the case in physiological conditions.

Using Eq~\eqref{eq_affinity_chemical_potentials} to circumvent the divergence issues is thus the first milestone towards a thermodynamically-consistent modelling of irreversible enzymatic reactions, as discussed in the next Section.

\section*{Results}
\label{section:results}

\subsection*{Nonequilibrium thermodynamics of the irreversible MM scheme}
\label{section_NET_MM}
We apply the concepts introduced in the previous Section and ref~\cite{wachtel_thermodynamically_2018} to a catalytic step proceeding according to the MM kinetics as if it were part of a biochemical pathway. To evaluate the steady-state EPR associated with the irreversible MM kinetics, the natural option is to approximate the reversible flux in Eq~\eqref{eq_EPR_rev_gen} by its irreversible counterpart:
\begin{equation}
\dot{\Sigma}^\mathrm{ss}_\mathrm{irr}=J_\mathrm{irr}\frac{\mathcal{A}_\mathrm{rev}}{T},
\label{eq_EPR_irr_gen}
\end{equation}
where $J_\mathrm{irr}$ is given by \eqref{eq_general_MM} and $\mathcal{A}_\mathrm{rev}$ is written as \eqref{eq_affinity_chemical_potentials} with positive and non-zero concentrations for S and P. The irreversible steady-state EPR can then be computed using the standard Gibbs free energies of formation of S and P and the kinetic parameters involved in $J_\mathrm{irr}$, \textit{i.e.} $K_M$ and $V$, which are typically available.

In order to check that Eq~\eqref{eq_EPR_irr_gen} is consistent with Eq~\eqref{eq_EPR_rev_gen}, we need to compare the EPR obtained in both cases. However, the comparison is only possible when the kinetic parameters of $J_\mathrm{rev}$ ($k_s$, $k_p$, $K_{Ms}$ and $K_{Mp}$) are also available, which is usually not the case. A more general and systematic analysis can be performed if the steady-state EPR is written in terms of the kinetic constants $\left\lbrace k_1, k_2, k_{-1}, k_{-2}\right\rbrace$, where $\left\lbrace k_1, k_2, k_{-1}\right\rbrace$ correspond to the available $K_M$ and $V$, while $k_{-2}$ is set to a positive and arbitrarily small value to mimic irreversibility. $k_{-2}$ can then be gradually increased to tend to a reversible reaction. Eq~\eqref{eq_EPR_rev_gen} becomes
\begin{equation}
\dot{\Sigma}^{ss}_{\mathrm{rev}}=\frac{\left[\mathrm{E}\right]_0\left(k_1\,k_2 \left[\mathrm{S}\right]-k_{-1}\,k_{-2} \left[\mathrm{P}\right]\right)}{k_1 \left[\mathrm{S}\right]+k_{-2} \left[\mathrm{P}\right]+k_{-1}+k_2} R \ln \frac{k_1\, k_2\, \left[\mathrm{S}\right]}{k_{-1}\, k_{-2}\,\left[\mathrm{P}\right]},
\label{eq_EPRss_rev}
\end{equation}
and Eq~\eqref{eq_EPR_irr_gen}
\begin{equation}
\dot{\Sigma}^{ss}_{\mathrm{irr}}=\frac{k_1 k_2 \left[\mathrm{E}\right]_0 \left[\mathrm{S}\right]}{k_1 \left[\mathrm{S}\right]+k_{-1}+k_2} R \ln \frac{k_1\, k_2\, \left[\mathrm{S}\right]}{k_{-1}\, k_{-2}\,\left[\mathrm{P}\right]}.
\label{eq_EPRss_irr_1}
\end{equation}
In Eq~\eqref{eq_EPRss_rev}, the flux and the force terms have always the same sign and the EPR is thus always positive. However, in Eq~\eqref{eq_EPRss_irr_1}, $J_\mathrm{irr}$ is always positive while $\mathcal{A}_\mathrm{rev}$ can take negative values if $k_1\, k_2\, \left[\mathrm{S}\right] < k_{-1}\, k_{-2}\,\left[\mathrm{P}\right]$, which leads to a negative EPR, in contradiction with the second law of thermodynamics. We can thus expect that Eq~\eqref{eq_EPRss_irr_1} approximates Eq~\eqref{eq_EPRss_rev} in a thermodynamically-consistent way when $k_{-2}$ and $\left[\mathrm{P}\right]$ are small enough, but how can we define a threshold? 

To investigate the conditions under which the use of the irreversible MM equation becomes problematic, we plot $\dot{\Sigma}^{ss}_{\mathrm{irr}}$ as a function of $\left[\mathrm{S}\right]$ and $\left[\mathrm{P}\right]$. The free parameters are the kinetic constants $\left\lbrace k_1, k_2, k_{-1}, k_{-2}\right\rbrace$, which are chosen to be of the same order of magnitude as the typical values of $K_M$ and $V$ reported in the literature. We also verify that the reaction is spontaneous in standard conditions, \textit{i.e.}\ $\Delta_{\mathrm{rev}} G^\circ<0$. Our results are independent of the enzyme total concentration ($\left[\mathrm{E}\right]_0$), which is set to 1 $\mu$M, and the temperature is set to 298.15 K to match the conditions of the thermodynamic tables used in the following.

\begin{figure}[t]
\centering
\includegraphics[width=\textwidth]{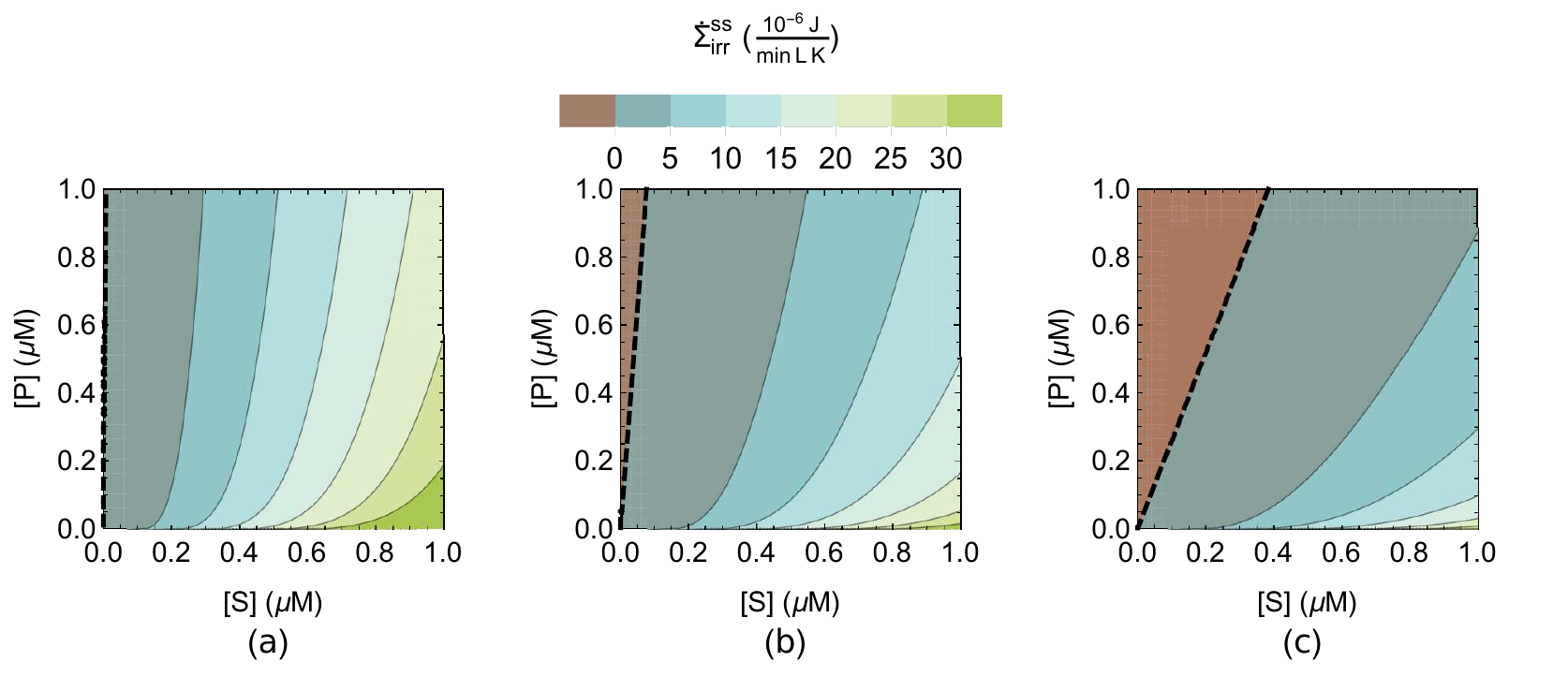}
\caption{{\bf Contour plot of the irreversible steady-state EPR as a function of substrate and product concentrations.} $k_{-2}$ is 0.01$\, \mu\mathrm{M}^{-1}$ min\textsuperscript{-1} in (a), $0.10\, \mu\mathrm{M}^{-1}$ min\textsuperscript{-1} in (b) and $0.50\, \mu\mathrm{M}^{-1}$ min\textsuperscript{-1} in (c), respectively. The other parameter values are $k_1=1\, \mu\mathrm{M}^{-1}$ min\textsuperscript{-1}, $k_{-1}=20$ min\textsuperscript{-1} and $k_{2}=26$ min\textsuperscript{-1}. Such values correspond to $K_M=46\, \mu$M and $V=26\, \mu$M min\textsuperscript{-1} and are for example of the same order of magnitude as the kinetic parameters reported for the tyrosine hydroxylase in the synthesis of dopamine from L-tyrosine \cite{nakashima_dopamine_1999}. The dashed black line corresponds to $\dot{\Sigma}^{\mathrm{ss}}_{\mathrm{irr}}=0$.}
\label{fig_EPRss}
\end{figure}

As shown in Fig~\ref{fig_EPRss}, $\dot{\Sigma}^{ss}_{\mathrm{irr}}<0$ for a certain range of concentration of the chemostatted species S and P, but this behaviour tends to disappear as $k_{-2}$ decreases.
More importantly, the boundary defining $\dot{\Sigma}^{ss}_{\mathrm{irr}}=0$ is always a straight line of equation
\begin{equation}
\left[\mathrm{P}\right]=\left[\mathrm{S}\right] \exp \left(-\frac{\Delta_{\mathrm{rev}} G^\circ}{R T}\right),
\label{eq_boundary}
\end{equation}
which can easily be derived by introducing $\Delta_{\mathrm{rev}} G^\circ$ into the last factor of Eq~\eqref{eq_EPRss_irr_1} \textit{via} $\Delta_{\mathrm{rev}} G^\circ=- R T \ln \frac{k_1\, k_2}{k_{-1}\, k_{-2}}$, where we assume the system satisfies the principle of detailed balance and the hypothesis of local equilibrium, and equalling the resulting equation to zero.

\begin{figure}[t!]
\centering
\includegraphics[width=0.4\textwidth]{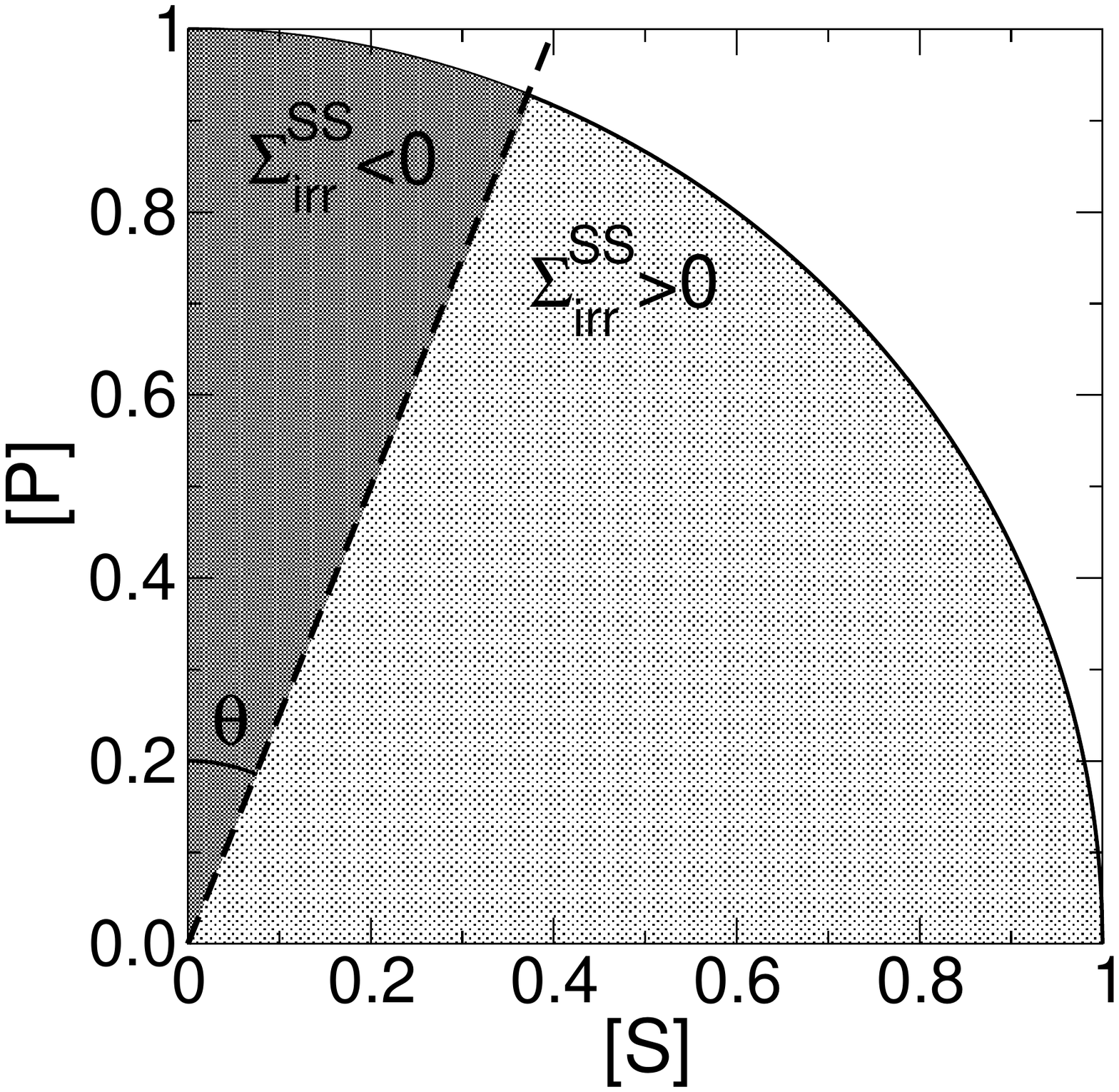}
\caption{{\bf Schematic representation of the areas of the concentration space taken into account in the calculation of $\mathbf{\xi}$.} The dark gray area corresponds to the sector where the steady-state EPR takes unphysical values.}
\label{fig_trigono}
\end{figure}

While the slope of \eqref{eq_boundary} and the area occupied by the region $\dot{\Sigma}^{\mathrm{ss}}_{\mathrm{irr}}<0$ in the $\left[\mathrm{S}\right]$-$\left[\mathrm{P}\right]$ plane (Fig~\ref{fig_EPRss}) already indicate the tendency of the system to generate unphysical thermodynamic results, a bounded quantity is more suitable to ease the comparison with other systems. We thus define a scaled coefficient, $\xi$, as the area of the sector determined by the angle $\theta$ in the first quadrant of the unit circle, \textit{i.e}.\ $\frac{\pi}{2 \pi} \theta$, normalised by the total area of the first quadrant, \textit{i.e}.\ $\frac{\pi}{4}$ (see Fig~\ref{fig_trigono}):
\begin{equation}
\xi=\frac{2 \theta}{\pi}.
\end{equation}
Since $0\leq\theta\leq \pi / 2$, $\xi$ is bounded between 0 and 1.
Additionally, there exists a direct relation between the slope of \eqref{eq_boundary} and the angle $\theta$
\begin{equation}
\tan\left(\frac{\pi}{2}-\theta\right)=\frac{1}{\tan\theta}=\exp \left(-\frac{\Delta_{\mathrm{rev}} G^\circ}{R T}\right),
\end{equation}
so that we can express $\xi$ as a function of the thermodynamic properties of the system
\begin{equation}
\xi=\frac{2}{\pi}\arctan\left[\exp\left(\frac{\Delta_{\mathrm{rev}} G^\circ}{R T}\right)\right]=\frac{2}{\pi}\arctan\left[\frac{1}{K}\right].
\label{eq_xi}
\end{equation}

We recall that, as a result of increasing efforts made to collect thermodynamic properties of biochemical compounds such as standard Gibbs free energy of formation ($\Delta_f G^\circ_{\sigma}$) \cite{noor_integrated_2012, flamholz_equilibratorbiochemical_2012, noor_consistent_2013}, $\Delta_\mathrm{rev} G^\circ$, and hence $\xi$, can in practice be computed for a wide range of reactions \textit{via} $\Delta_\mathrm{rev} G^\circ=\Delta_f G^\circ_\mathrm{P}-\Delta_f G^\circ_\mathrm{S}$ for the catalytic step $\mathrm{E}+\mathrm{S} \rightleftharpoons \mathrm{ES}\rightarrow \mathrm{E} + \mathrm{P}$.

%

\subsection*{Rule of thumb}
\label{section_rule}
\begin{figure}[t!]
\centering
\includegraphics[width=0.4\textwidth]{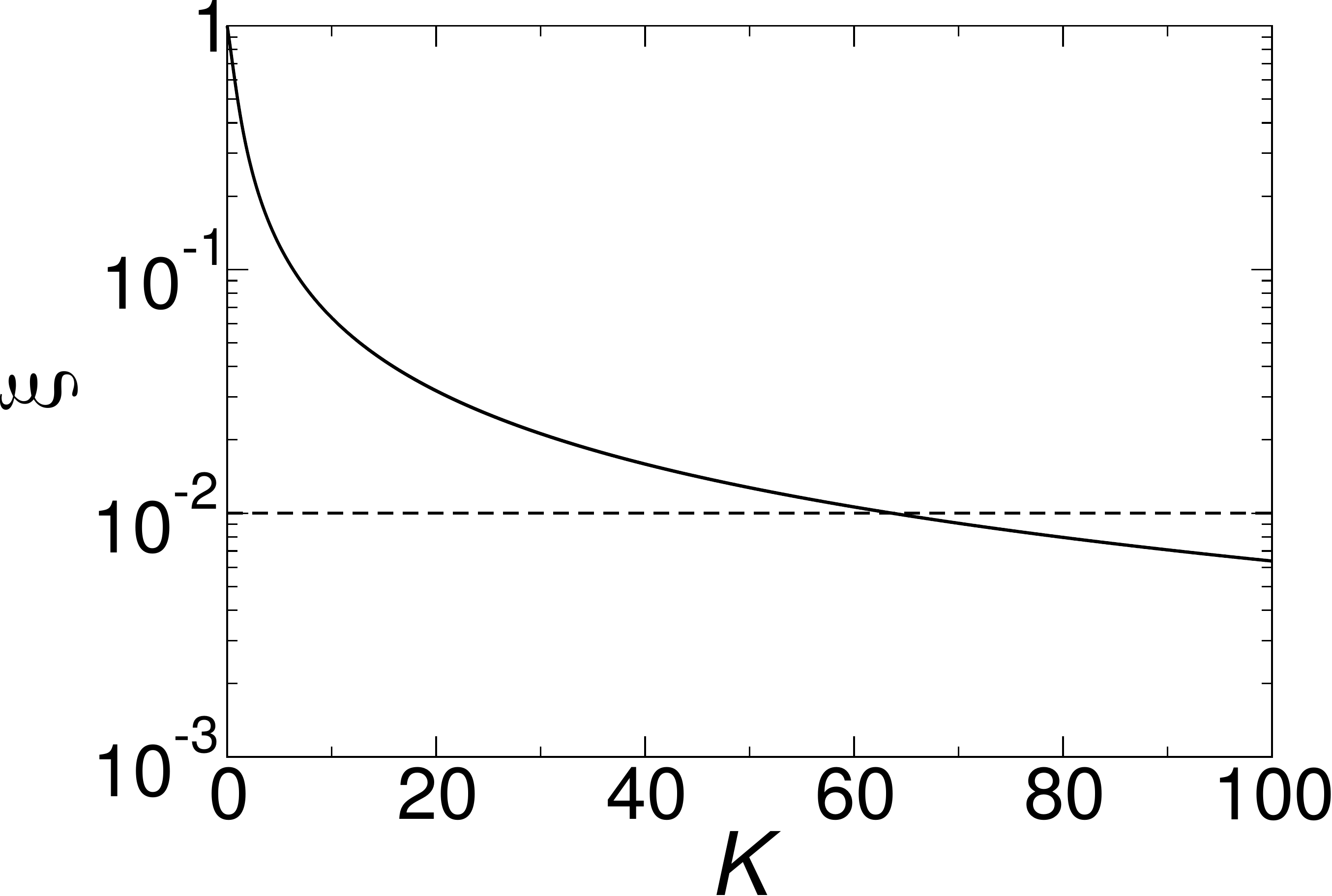}
\caption{{\bf Plot of $\xi$ as a function of the equilibrium constant $K$.} The dashed line at $\xi=0.01$ corresponds to the threshold below which the irreversible MM equation can be used without compromising the thermodynamic relevance of the model.}
\label{fig_xi_vs_K}
\end{figure}

Although Eq~\eqref{eq_xi} provides a scale to compare different biochemical reactions, the threshold between reversible and irreversible reactions still needs to be estimated. If $\xi$ is plotted as a function of $K$ (Fig~\ref{fig_xi_vs_K}) and the threshold is set to a value of 1\%, the approximate value of $K$ above which we can consider the reaction as irreversible is around 60. This value corresponds to $\Delta_{\mathrm{rev}} G^\circ \approx -10$ kJ/mol at 298.15 K. Note that for the conditions of Fig~\ref{fig_EPRss}, $\xi$ is equal to 0.49\% in (a), 4.9\% in (b) and  23\% in (c), which means that the irreversible MM equation can be applied, according to our criterion, in conditions (a) only.

As a test, we applied this criterion to the thermodynamic data available for the Krebs cycle and glycolysis. Biochemical reactions are usually assumed to proceed at given pH and ionic strength, although protons and other ions can be consumed or produced during the reaction. Apparent equilibrium constants ($K'$) and standard transformed Gibbs free energies of reaction ($\Delta_\mathrm{rev} G^\circ\,'$) are thus calculated \cite{alberty_thermodynamics_2003}. For example, a reaction $\mathrm{HA} \rightleftharpoons \mathrm{A}^- + \mathrm{H}^+$ whose equilibrium constant is $K=\frac{\left[\mathrm{H}^+\right] \left[\mathrm{A}^-\right]}{\left[\mathrm{HA}\right]}$ is characterised by
\begin{eqnarray}
\Delta_\mathrm{rev} G^\circ\,'&=&-RT \ln K'\\
&=& -RT \ln K + RT \ln \left[\mathrm{H}^{+}\right]\\
&=& \Delta_\mathrm{rev} G^\circ + RT \ln \left[\mathrm{H}^{+}\right].
\end{eqnarray}
For the Krebs cycle and glycolysis, reactions take place in the mitochondrial matrix (pH$\approx$8) and in the cytosol (pH$\approx$7), respectively, so we used the standard transformed free Gibbs energies of reaction at appropriate pH. We spotted 3 reactions with $\Delta_{\mathrm{rev}} G^\circ\,' \leq -10$ kJ/mol in the Krebs cycle, namely the condensation of acetyl-CoA and oxaloacetate to form citrate and the oxidative decarboxylations of isocitrate and $\alpha$-ketoglutarate. As for the glycolytic pathway, the conversions of glucose into glucose-6-phosphate, of fructose-6-phosphate into fructose-1,6-bisphosphate and of phosphoenolpyruvate into pyruvate are also satisfying $\xi<1\%$. In both examples, applying our heuristic criterion leads to conclusions in agreement with biochemistry textbooks, where the aforementioned reactions are characterised as irreversible given their strongly negative $\Delta_\mathrm{rev} G$ at physiological concentrations of metabolites~\cite{nelson_lehninger_2008}.

\subsection*{Comparison between the reversible and irreversible EPR at steady state}
\label{section_comparison}
The aforementioned criterion can be seen by modellers as a qualitative indicator of whether the irreversible MM rate is a thermodynamically acceptable approximation. However, it provides no information regarding the quantitative agreement between the reversible and irreversible steady-state EPR in the valid domain, which we now examine in more detail.

\begin{figure}[t!]
\centering
\includegraphics[width=\textwidth]{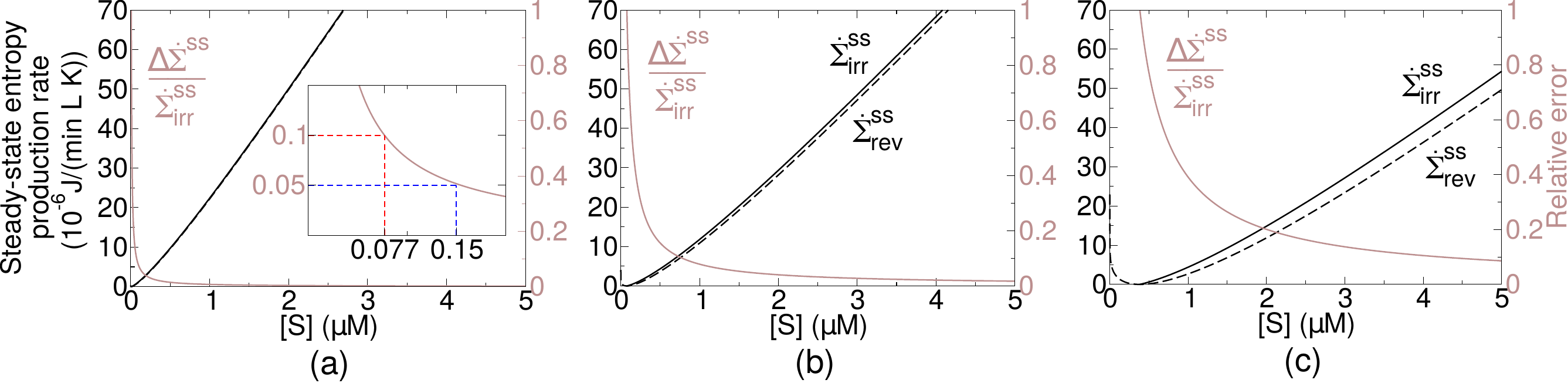}
\caption{{\bf Comparison and relative error between $\mathbf{\dot{\Sigma}_{\mathrm{rev}}^{\mathrm{ss}}}$ and $\mathbf{\dot{\Sigma}_{\mathrm{irr}}^{\mathrm{ss}}}$.} $k_{-2}$ is equal to $0.01\, \mu\mathrm{M}^{-1}$ min\textsuperscript{-1} in (a), to $0.10\, \mu\mathrm{M}^{-1}$ min\textsuperscript{-1} in (b) and to $0.50\, \mu\mathrm{M}^{-1}$ min\textsuperscript{-1} in (c). The inset in panel (a) is a zoom on the relative error curve. $\left[\mathrm{P}\right]$ is equal to $1\, \mu$M. $\dot{\Sigma}_{\mathrm{rev}}^{\mathrm{ss}}$ and $\dot{\Sigma}_{\mathrm{irr}}^{\mathrm{ss}}$ are represented by the dashed and plain black lines, respectively, while the relative error, defined as $\frac{\Delta \dot{\Sigma}^{\mathrm{ss}}}{\dot{\Sigma}_{\mathrm{irr}}^{\mathrm{ss}}}=\frac{\dot{\Sigma}_{\mathrm{irr}}^{\mathrm{ss}}-\dot{\Sigma}_{\mathrm{rev}}^{\mathrm{ss}}}{\dot{\Sigma}_{\mathrm{irr}}^{\mathrm{ss}}}$, is shown in light brown.}
\label{fig_comparison_error}
\end{figure}

We define the absolute error $\Delta \dot{\Sigma}^{\mathrm{ss}}$ as 
\begin{equation}
\Delta \dot{\Sigma}^{\mathrm{ss}}=\dot{\Sigma}^{\mathrm{ss}}_{\mathrm{irr}}-\dot{\Sigma}^{\mathrm{ss}}_{\mathrm{rev}}=\left(J_{\mathrm{irr}}-J_{\mathrm{rev}}\right) \mathcal{A}_\mathrm{rev}.
\end{equation}
Since $J_{\mathrm{irr}}\geq J_{\mathrm{rev}}$, the sign of $\Delta \dot{\Sigma}^{\mathrm{ss}}$ depends only on $\mathcal{A}_\mathrm{rev}$, as is the case for $\dot{\Sigma}^{\mathrm{ss}}_{\mathrm{irr}}$, and is thus positive in the \textit{physical domain} corresponding to $\dot{\Sigma}^{\mathrm{ss}}_{\mathrm{irr}}>0$. By definition, $\Delta \dot{\Sigma}^{\mathrm{ss}}$ is zero at equilibrium and tends to zero as $\left[\mathrm{S}\right]\rightarrow \infty$, so it exhibits a maximum for an intermediate value of $\left[\mathrm{S}\right]$. This absolute error can be normalised to obtain the relative error
\begin{equation}
\frac{\Delta \dot{\Sigma}^{\mathrm{ss}}}{\dot{\Sigma}^{\mathrm{ss}}_{\mathrm{irr}}}=1-\frac{J_{\mathrm{rev}}}{J_\mathrm{irr}}=1-\frac{\left(k_1 k_2 \left[\mathrm{S}\right]-k_{-1} k_{-2} \left[\mathrm{P}\right]\right) \left(k_1 \left[\mathrm{S}\right] +k_{-1}+k_2\right)}{k_1 k_2 \left[\mathrm{S}\right] \left(k_1 \left[\mathrm{S}\right]+k_2 \left[\mathrm{P}\right]+k_{-1}+k_2\right)},
\end{equation}
which is bounded between 0 in 1 since $0 \leq J_{\mathrm{rev}}\leq J_{\mathrm{irr}}$ in the physical domain. In Fig~\ref{fig_comparison_error}, we compare the steady-state EPR in the reversible and irreversible cases. The relative error is equal to 1 at equilibrium ($\dot{\Sigma}^{\mathrm{ss}}_{\mathrm{irr}}=\dot{\Sigma}^{\mathrm{ss}}_{\mathrm{rev}}$=0) before monotonously decreasing towards zero as $\left[\mathrm{S}\right]$ is increased.
Also, the convergence of the relative error to zero becomes slower as $k_{-2}$ (Figures \ref{fig_comparison_error}.a to \ref{fig_comparison_error}.c) and $\left[\mathrm{P}\right]$ (data not shown) are increased. This result is not surprising since these parameters are supposed to tend to zero in the irreversible limit. For very small values of $\left[\mathrm{S}\right]$, the concentration of P is comparatively high and the contribution of the reverse reaction to $\dot{\Sigma}^{\mathrm{ss}}_{\mathrm{rev}}$ is no longer negligible. As a result, the discrepancy between $\dot{\Sigma}^{\mathrm{ss}}_{\mathrm{irr}}$ and $\dot{\Sigma}^{\mathrm{ss}}_{\mathrm{rev}}$ is also larger.

The relative error for conditions of Fig~\ref{fig_comparison_error}.a, where our heuristic criterion is satisfied, drops below 10\% at $\left[\mathrm{S}\right]\approx 0.077\, \mu$M and below 5\% at $\left[\mathrm{S}\right]\approx 0.15\, \mu$M. For physiological values of $\left[\mathrm{S}\right]$, which are in the range of $170\, \mu$M for the reaction studied in the example of Fig~\ref{fig_EPRss} \cite{brodnik_l-tyrosine_2017}, the error is thus negligible. More generally, the irreversible MM equation should not be used to compute the steady-state EPR close to equilibrium. In that case, the reversible flux should be used instead.

\section*{Discussion}
We have derived a robust validity criterion for the irreversible MM equation, based on nonequilibrium thermodynamics. This criterion is particularly relevant for reactions proceeding in chemostatted environments, where the concentrations of S and P are steady, as it might be the case in biosynthesis, for example. When satisfied, it ensures a thermodynamically-consistent modelling of the enzymatic reaction. Moreover, the steady-state EPR can then be computed on the basis of the irreversible MM flux, leading to a negligible error with respect to the reversible case.

While the irreversibility of more complex mechanisms remains a question to be explored, our approach could nevertheless give us insights into MM reactions whose detailed kinetic properties are not yet available or difficult to obtain. Indeed, this criterion depends only on the standard Gibbs free energy of reaction, which can be computed from the -- usually available -- standard Gibbs free energies of formation of the species involved. In particular, our approach can be applied to a succession of $n$ irreversible MM steps of the form:
\begin{eqnarray}
\mathrm{E}_1 + \mathrm{S} \xrightleftharpoons[k_{-1}]{k_1} &\mathrm{E}_1\mathrm{S}& \xrightarrow{k_2} \mathrm{E}_1 + \mathrm{I}_1\nonumber\\
\mathrm{E}_2 + \mathrm{I}_1 \xrightleftharpoons[k_{-3}]{k_3} &\mathrm{E}_2\mathrm{I}_1& \xrightarrow{k_4} \mathrm{E}_2 + \mathrm{I}_2\nonumber\\
&...&\nonumber\\
\mathrm{E}_n + \mathrm{I}_{n-1} \xrightleftharpoons[k_{-\left(2n-1\right)}]{k_{\left(2n-1\right)}} &\mathrm{E}_n\mathrm{I}_{n-1}& \xrightarrow{k_{2n}} \mathrm{E}_n + \mathrm{P}\nonumber,
\end{eqnarray}
where S and P are chemostatted. The procedure would start by checking, for each reaction, if the criterion for irreversibility is satisfied, based on a estimation of the steady-state concentrations of S, $\left\lbrace I_i\right\rbrace$ and P, which could be provided experimentally. If so, the steady-state flux associated with the resulting global reaction $\mathrm{S} \rightarrow \mathrm{P}$ is given by the irreversible MM equation rate of the first catalytic step and the steady-state EPR by
\begin{eqnarray}
\dot{\Sigma}^\mathrm{ss}_\mathrm{irr}&=&\frac{k_1 k_2 \left[\mathrm{E}_1\right]_0 \left[\mathrm{S}\right]}{k_1 \left[\mathrm{S}\right] + k_{-1} + k_2}R\ln \frac{\prod\limits_i^{2n} k_i \left[\mathrm{S}\right]}{\prod\limits_i^{2n} k_{-i} \left[\mathrm{P}\right]}\\
&=& \frac{k_{\mathrm{cat},1}\left[\mathrm{E}_1\right]_0 \left[\mathrm{S}\right]}{\left[\mathrm{S}\right]+K_{M, 1}} \left[R \ln \frac{\left[\mathrm{S}\right]}{\left[\mathrm{P}\right]}-\frac{\left(\Delta_f G^\circ_\mathrm{P}-\Delta_f G^\circ_\mathrm{S}\right)}{T}\right],
\end{eqnarray}
where $\Delta_f G^\circ_\mathrm{S}$ and $\Delta_f G^\circ_\mathrm{P}$ are the standard Gibbs free energy of formation of compounds S and P, respectively. It is thus possible to calculate the steady-state EPR for a global reaction, whose independent catalytic steps are irreversible, only on the basis of experimentally available quantities such as $k_\mathrm{cat,1}$, $K_{M,1}$, the estimated concentrations of S, $\left[\mathrm{E}_1\right]_0$ and P, and the standard Gibbs free energies of formation of S and P. This approach might be particularly relevant to estimate the efficiency of metabolic pathways.

\section*{Acknowledgments}
VV is funded by the Complex Living Systems Initiative at the University of Luxembourg. FA and ME are funded by the European Research Council project NanoThermo (ERC-2015-CoG Agreement No. 681456).

\section*{Author Contributions}
FA and ME designed the study. VV performed the simulations and the data were analysed by FA and VV. VV wrote the paper. FA, ME and VV carried out the revision of the manuscript.

\nolinenumbers

%
%
%

%
%
%

\begin{thebibliography}{10}

\bibitem{murray_mathematical_2002}
Murray JD.
\newblock Mathematical {Biology}: {I}. {An} {Introduction}.
\newblock 3rd ed. Interdisciplinary {Applied} {Mathematics}. New York:
  Springer-Verlag; 2002.

\bibitem{keener_mathematical_2009}
Keener J, Sneyd J.
\newblock Mathematical {Physiology}: {I}: {Cellular} {Physiology}.
\newblock 2nd ed. Interdisciplinary {Applied} {Mathematics}. New York:
  Springer-Verlag; 2009.

\bibitem{lin_mathematics_1988}
Lin CC, Segel LA.
\newblock Mathematics {Applied} to {Deterministic} {Problems} in the {Natural}
  {Sciences}.
\newblock Classics in {Applied} {Mathematics}. Philadelphia: Society for
  Industrial and Applied Mathematics; 1988.

\bibitem{van_kampen_elimination_1985}
Van~Kampen NG.
\newblock Elimination of fast variables.
\newblock Phys Rep. 1985;124(2):69--160.

\bibitem{shiner_fluxes_1981}
Shiner JS.
\newblock Fluxes and the elimination of fast-relaxing variables.
\newblock J Stat Phys. 1981;26(3):555--565.

\bibitem{wachtel_thermodynamically_2018}
Wachtel A, Rao R, Esposito M.
\newblock Thermodynamically consistent coarse graining of biocatalysts beyond
  {Michaelis}–{Menten}.
\newblock New J Phys. 2018;20(4):042002.
\newblock doi:{10.1088/1367-2630/aab5c9}.

\bibitem{beard_chemical_2008}
Beard DA, Qian H.
\newblock Chemical {Biophysics}: {Quantitative} {Analysis} of {Cellular}
  {Systems}.
\newblock Cambridge {Texts} in {Biomedical} {Engineering}. Cambridge: Cambridge
  University Press; 2008.

\bibitem{cornish-bowden_fundamentals_2013}
Cornish-Bowden A.
\newblock Fundamentals of {Enzyme} {Kinetics}.
\newblock John Wiley \& Sons; 2013.

\bibitem{bisswanger_enzyme_2017}
Bisswanger H.
\newblock Enzyme {Kinetics}: {Principles} and {Methods}.
\newblock Weinheim: John Wiley \& Sons; 2017.

\bibitem{curien_understanding_2009}
Curien G, Bastien O, Robert-Genthon M, Cornish-Bowden A, Cárdenas ML, Dumas R.
\newblock Understanding the regulation of aspartate metabolism using a model
  based on measured kinetic parameters.
\newblock Mol Syst Biol. 2009;5:271.

\bibitem{min_when_2006}
Min W, Gopich IV, English BP, Kou SC, Xie XS, Szabo A.
\newblock When {Does} the {Michaelis}−{Menten} {Equation} {Hold} for
  {Fluctuating} {Enzymes}?
\newblock J Phys Chem B. 2006;110(41):20093--20097.

\bibitem{kolomeisky_michaelismenten_2011}
Kolomeisky AB.
\newblock Michaelis–{Menten} relations for complex enzymatic networks.
\newblock J Chem Phys. 2011;134(15):155101.

\bibitem{cao_michaelismenten_2011}
Cao J.
\newblock Michaelis−{Menten} {Equation} and {Detailed} {Balance} in
  {Enzymatic} {Networks}.
\newblock J Phys Chem B. 2011;115(18):5493--5498.

\bibitem{gorban_thermodynamics_2013}
Gorban AN, Mirkes EM, Yablonsky GS.
\newblock Thermodynamics in the limit of irreversible reactions.
\newblock Physica A. 2013;392(6):1318--1335.

\bibitem{segel_quasi-steady-state_1989}
Segel LA, Slemrod M.
\newblock The {Quasi}-{Steady}-{State} {Assumption}: {A} {Case} {Study} in
  {Perturbation}.
\newblock SIAM Rev. 1989;31(3):446--477.

\bibitem{lee_kinetics_2000}
Lee HS, Hong J.
\newblock Kinetics of glucose isomerization to fructose by immobilized glucose
  isomerase: anomeric reactivity of d-glucose in kinetic model.
\newblock J Biotechnol. 2000;84(2):145--153.

\bibitem{rao_nonequilibrium_2016}
Rao R, Esposito M.
\newblock Nonequilibrium {Thermodynamics} of {Chemical} {Reaction} {Networks}:
  {Wisdom} from {Stochastic} {Thermodynamics}.
\newblock Phys Rev X. 2016;6(4):041064.

\bibitem{de_groot_non-equilibrium_1984}
De~Groot SR, Mazur P.
\newblock Non-{Equilibrium} {Thermodynamics}.
\newblock New York: Dover; 1984.

\bibitem{nakashima_dopamine_1999}
Nakashima A, Mori K, Suzuki T, Kurita H, Otani M, Nagatsu T, et~al.
\newblock Dopamine {Inhibition} of {Human} {Tyrosine} {Hydroxylase} {Type} 1
  {Is} {Controlled} by the {Specific} {Portion} in the {N}-{Terminus} of the
  {Enzyme}.
\newblock J Neurochem. 1999;72(5):2145--2153.

\bibitem{noor_integrated_2012}
Noor E, Bar-Even A, Flamholz A, Lubling Y, Davidi D, Milo R.
\newblock An integrated open framework for thermodynamics of reactions that
  combines accuracy and coverage.
\newblock Bioinformatics. 2012;28(15):2037--2044.

\bibitem{flamholz_equilibratorbiochemical_2012}
Flamholz A, Noor E, Bar-Even A, Milo R.
\newblock {eQuilibrator}—the biochemical thermodynamics calculator.
\newblock Nucleic Acids Res. 2012;40(Database issue):D770--D775.

\bibitem{noor_consistent_2013}
Noor E, Haraldsdóttir HS, Milo R, Fleming RMT.
\newblock Consistent {Estimation} of {Gibbs} {Energy} {Using} {Component}
  {Contributions}.
\newblock PLoS Comput Biol. 2013;9(7):e1003098.

\bibitem{alberty_thermodynamics_2003}
Alberty RA.
\newblock Thermodynamics of {Biochemical} {Reactions}.
\newblock Hoboken, New Jersey: John Wiley \& Sons; 2003.

\bibitem{nelson_lehninger_2008}
Nelson DL, Lehninger AL, Cox MM.
\newblock Lehninger {Principles} of {Biochemistry}.
\newblock 5th ed. New York: W. H. Freeman; 2008.

\bibitem{brodnik_l-tyrosine_2017}
Brodnik ZD, Double M, España RA, Jaskiw GE.
\newblock L-{Tyrosine} {Availability} {Affects} {Basal} and {Stimulated}
  {Catecholamine} {Indices} in {Prefrontal} {Cortex} and {Striatum} of the
  {Rat}.
\newblock Neuropharmacology. 2017;123:159--174.

\end{thebibliography}

\end{document}